\documentclass[]{pasj01}
\draft

\Received{}
\Accepted{}
 
\begin{document} 

\title{ 
Silicon and Strontium abundances of very metal-poor stars determined from near-infrared spectra }

\author{Wako \textsc{Aoki}\altaffilmark{1,2}}
\altaffiltext{1}{National Astronomical Observatory, 
 2-21-1 Osawa, Mitaka, Tokyo 181-8588, Japan }
\altaffiltext{2}{Department of Astronomical Science, School of Physical Sciences
, The Graduate University of Advanced Studies (SOKENDAI), 2-21-1 Osawa, Mitaka,
Tokyo 181-8588, Japan}
\email{aoki.wako@nao.ac.jp}

\author{Timothy C. \textsc{Beers}\altaffilmark{3}}
\altaffiltext{3}{Department of Physics and JINA Center for the Evolution of the Elements, University of Notre Dame, Notre Dame, IN 46556, USA}
\email{timothy.c.beers.5@nd.edu}

\author{Satoshi \textsc{Honda}\altaffilmark{4}}
\altaffiltext{4}{Nishi-Harima Astronomical Observatory, Center for Astronomy,
University of Hyogo, 407-2 Nishigaichi, Sayo-cho, Sayo, Hyogo 679-5313, Japan}
\email{honda@nhao.jp}

\author{Hiroyuki T. \textsc{Ishikawa}\altaffilmark{5}}
\altaffiltext{5}{Astrobiology Center, 2-21-1 Osawa, Mitaka, Tokyo 181-8588, Japan}

\author{Tadafumi \textsc{Matsuno}\altaffilmark{6}}
\altaffiltext{6}{Kapteyn Astronomical Institute, University of Groningen  \\ Lan
dleven 12, 9747 AD Groningen, The Netherlands}
\email{matsuno@astro.rug.nl}

\author{Vinicius M. \textsc{Placco}\altaffilmark{7}}
\altaffiltext{7}{NSF’s NOIRLab, 950 N. Cherry Ave., Tucson, AZ 85719, USA}
\email{vinicius.placco@noirlab.edu}

\author{Jinmi \textsc{Yoon}\altaffilmark{3,8}}
\altaffiltext{8}{Space Telescope Science Institute, 3700 San Martin Dr., Baltimore, MD 21218, USA}
\email{jyoon@stsci.edu}

\author{Hiroki   \textsc{Harakawa}\altaffilmark{9}}  
\altaffiltext{9}{Subaru Telescope, 650 N. Aohoku Place, Hilo, HI 96720, USA}
\email{harakawa@naoj.org}

 
\author{Teruyuki \textsc{Hirano}\altaffilmark{1,2,5}}  

\author{Klaus \textsc{Hodapp}\altaffilmark{10}}
\altaffiltext{10}{University of Hawaii, Institute for Astronomy, 640 N. Aohoku Place, Hilo, HI 96720, USA}
\email{hodapp@hawaii.edu}

\author{Masato  \textsc{Ishizuka}\altaffilmark{11}}  
\altaffiltext{11}{Department of Astronomy, Graduate School of Science, The University of Tokyo, 7-3-1 Hongo, Bunkyo-ku, Tokyo 113-0033, Japan}
\email{ishizuka@astron.s.u-tokyo.ac.jp}

\author{Shane  \textsc{Jacobson}\altaffilmark{10}}
\email{shanemj@hawaii.edu}


\author{Takayuki   \textsc{Kotani}\altaffilmark{1,5,2}}  
\email{t.kotani@nao.ac.jp}

\author{Tomoyuki   \textsc{Kudo}\altaffilmark{8}}  
\email{kudotm@subaru.naoj.org}

\author{Takashi   \textsc{Kurokawa}\altaffilmark{5,13}}    
\altaffiltext{14}{Institute of Engineering, Tokyo University of Agriculture and Technology, 2-24-16, Nakacho, Koganei, Tokyo, 184-8588, Japan}
\email{tkuro@cc.tuat.ac.jp}

\author{Masayuki   \textsc{Kuzuhara}\altaffilmark{1,5}}  
\email{m.kuzuhara@nao.ac.jp}

\author{Jun   \textsc{Nishikawa}\altaffilmark{1,2,5}}  
\email{jun.nishikawa@nao.ac.jp}

\author{Masashi   \textsc{Omiya}\altaffilmark{1,5}}  
\email{omiya.masashi@nao.ac.jp}

\author{Takuma  \textsc{Serizawa}\altaffilmark{1,13}}    
\email{serizawa@go.tuat.ac.jp}

\author{Motohide  \textsc{Tamura}\altaffilmark{1,5,11}}      
\email{motohide.tamura@nao.ac.jp}

\author{Akitoshi   \textsc{Ueda}\altaffilmark{1,2,5}}   
\email{a.ueda@nao.ac.jp}

\author{Sébastien  \textsc{Vievard}\altaffilmark{8}}
\email{vievard@naoj.org}


\KeyWords{nuclear reactions, nucleosynthesis, abundances --- stars:abundances --- stars: Population II}

\maketitle

\begin{abstract}
Silicon and Strontium are key elements to explore the nucleosynthesis and chemical evolution of the Galaxy by measurements of very metal-poor stars. There are, however, only a few useful spectral lines of these elements in the optical range that are measurable for such low-metallicity stars. Here we report on abundances of these two elements determined from near-infrared high-resolution spectra obtained with the Subaru Telescope Infrared Doppler instrument (IRD). Si abundances are determined for as many as 26 Si lines for six very and extremely metal-poor stars ($-4.0<$ [Fe/H]$<-1.5$), which significantly improves the reliability of the abundance measurements. All six stars, including three carbon-enhanced objects, show over-abundances of Si ([Si/Fe]$\sim +0.5$). Two stars with [Fe/H]$\sim -1.5$ have relatively small over-abundances. The [Mg/Si] ratios agree with the solar value, except for one metal-poor star with carbon excess. Strontium abundances are determined from the triplet lines for four stars, including two for the first time. The consistency of the Sr abundances determined from near-infrared and optical spectra require further examination from additional observations.

\end{abstract}

\section{Introduction}

Low-mass stars with very low metallicity found in the Milky Way 
are believed to have formed in the very early stage of chemical
evolution, reflecting the products of the first and early generations of
massive stars and supernova explosions (e.g., \cite{Nomoto2013ARAA}). 
Observational studies of the elemental abundances for very
metal-poor (VMP: [Fe/H]$<-2$)\footnote{[A/B]
  = $\log(N_{\rm A}/N_{\rm B}) -\log(N_{\rm A}/N_{\rm B})_{\odot}$,
  and $\log\epsilon_{\rm A} =\log(N_{\rm A}/N_{\rm H})+12$ for
  elements A and B.}  stars play unique roles to
constrain the nucleosynthesis processes and the characteristics of their
progenitor stars in the early universe. The most useful abundance
ratios are those between the $\alpha$-elements and iron, which reflect the
masses of the progenitors of core-collapse supernovae (e.g., \cite{Heger2010ApJ, Ishigaki2018ApJ}) as well as the contributions from type-Ia
supernovae: these provide useful constraints on chemical-evolution models and formation scenarios of the Milky Way halo, including the accretion of dwarf galaxies. Neutron-capture elements are also important as records of explosive events
such as neutron star mergers and exotic supernovae \citep{Kajino2019PrPNP, Cowan2021RvMP}.

Among the $\alpha$-elements, Si, as well as Mg, are the most abundant ($\log\epsilon$(Si)$=7.51$ and $\log\epsilon$(Mg)$=7.60$; \cite{Asplund2009ARAA}) and are a
key for studying early chemical enrichment. Silicon is also a major source of dust grains that play crucial roles in star formation and stellar mass loss. There are, however, only a
few Si spectral lines in the optical range that are useful to
determine Si abundances in VMP stars, whereas Mg abundances are studied based on several lines in the optical range with a variety of strengths. In particular, for extremely
metal-poor (EMP: [Fe/H]$<-3$) stars, most of the Si abundance results
reported to date (e.g., \cite{Cayrel2004AA,Yong2013ApJ}) rely on only 
two lines in the blue range (390.5 and 410.3nm), which in low-mass
metal-poor stars are usually weak, and where spectrometers are less
efficient.  As a result, Mg and Ca are more frequently used to
represent the $\alpha$-elements. However, Si should be investigated as a
major product during both massive star evolution and supernovae
explosions, whereas Mg and Ca mostly represent the products in massive star evolution and supernova explosion, respectively. Standard models of nucleosynthesis and chemical evolution
do not predict large scatter in the abundance ratios of [Si/Fe], and special mechanisms would be 
required to explain outliers.
Thus, more reliable Si abundances based on larger numbers of spectral
lines are required to examine the abundance scatter and to identify the presence of outliers, if any.

Strontium is also a key element to constrain neutron-capture processes in the
early Galaxy. Many processes and sites are proposed for Sr production: the (main)
r-process, the weak-r process or Lighter Elements Primary Process (LEPP)  \citep{Wanajo2006NuPhA, Travaglio2004ApJ}, the main s-process in the case of
objects affected by mass transfer from Asymptotic Giant Branch (AGB) stars in binary systems (CEMP-s, or CH
stars), and the weak s-process in massive stars (e.g., \cite{Kappeler2011RvMP}). There are two resonance lines in the blue range, which are
very useful to determine Sr abundances in EMP stars. The lines are,
however, too strong to determine accurate abundances in stars
with relatively high metallicity ([Fe/H]$\gtrsim -2$) or with excesses
of Sr. Unfortunately, there are no other useful Sr lines with moderate
strengths in the optical range. As a result, Sr abundances are less
certain than abundances of another key neutron-capture element, Ba,
which has weaker lines in the red spectral range.
 
These two elements, Si and Sr, both have useful spectral lines in the near-infrared
range. There are many Si lines with a variety of strengths that are
detectable in red giants in the $Y$-, $J$- and $H$-bands, even for stars
with low metallicity. Si abundances are studied based on $H$-band spectra by APOGEE \citep{Jonsson2020AJ}. Most of the targets are disk stars, but some metal-poor stars with [Fe/H]$\lesssim -2$ are also covered. This demonstrates that near-infrared spectra with higher resolution are useful to study Si abundances in VMP and EMP stars. 
There are triplet lines of Sr in the $Y$-band,
which are detectable in EMP red giants, but they are not as strong as
those of the resonance lines in the blue region.

We here report on abundance analyses of these lines to obtain reliable Si and Sr abundances for six metal-poor stars with [Fe/H] from $-4$ to $-1.5$. Our near-infrared observations are reported in Section~\ref{sec:obs}. Section~\ref{sec:ana} provides details of the abundance analyses and error estimates. The Si abundance results and detection limits for future studies are discussed in Section~\ref{sec:disc}

\section{Observations}\label{sec:obs}
The near-infrared spectra were obtained with the Subaru Telescope InfraRed Doppler instrument (IRD; \cite{Tamura2012SPIE, Kotani2018SPIE}) on July 25, 2020 (UT). The spectra cover the $Y$-, $J$- and $H$-bands with spectral resolution of $R\sim70,000$. One pixel corresponds to about 6~pm at around 1~$\mu$m, resulting in about 2.4~pixel sampling of the resolution element.

The objects studied to determine Si and Sr abundances are listed in table~\ref{tab:param}. They are metal-poor stars that have been well-studied by previous work to determine elemental abundances from optical spectra. HD~221170, HD~4306, and LAMOST~J~2217+2104 are metal-poor red giants with a variety of [Fe/H] values from $-3.9$ to $-2.2$. LAMOST~J~2217+2104 is a carbon-enhanced star with excesses of Mg and Si \citep{Aoki2018PASJ}. BD+44$^{\circ}$493 is an extremely metal-poor ([Fe/H]$=-3.8$) subgiant star with carbon excess \citep{Ito2009ApJL}. HD~201626 is a very metal-poor CH star showing large excesses of carbon and heavy neutron-capture elements. The variation of radial velocities of this object \citep{McClure1990ApJ}, as well as the abundance pattern, indicates that this star was affected by mass accretion from the companion in a binary system when it was an AGB star \citep{Vaneck2003AA, Placco2015ApJ}. For this star, many weak Si lines in the optical range have been measured by \citet{Placco2015ApJ}, due to the relatively high metallicity ([Fe/H]$=-1.5$) and low temperature of this object. HD~25329 is a cool main-sequence star with [Fe/H]$=-1.6$ \citep{Luck2017AJ}. The lines of main-sequence stars are weaker than giants in general because of the larger continuous opacity of H$^{-}$ in cool main-sequence stars. However, the Si and Sr lines in the near-infrared range are detectable in cool main-sequence stars with this metallicity. 

Data reduction of the IRD spectra was conducted using the pipeline based on PyRAF, which adopts the data processes reported in \citet{Kuzuhara2018SPIE} and Kuzuhara et al. (in preparation). The procedure includes bias correction, removal of
correlated read-out noise, and extracting spectra for stellar and
calibration data by tracing spectra on 2D images using flat-fielding
images. The wavelength calibrations of the extracted stellar spectra are made by comparing the Th-Ar spectra obtained in our program with the reference Th-Ar spectra. The wavelengths of the reference spectra have been carefully calibrated by the IRD team based on the Th-Ar atlas of \citet{Kerber2008ApJS} and the spectra of laser frequency comb \citep{Hirano2020PASJ}.

Telluric absorption lines are identified by comparing the spectra of bright metal-poor stars in our sample. The lines that show no wavelength shift for all spectra, regardless their radial velocities, are treated as telluric lines. Stellar spectral lines that are not affected by telluric lines are selected for the abundance analysis in the present work. This treatment does not significantly reduce the number of available lines for abundance analyses.

\begin{figure}
 \begin{center}
   \includegraphics[width=13cm]{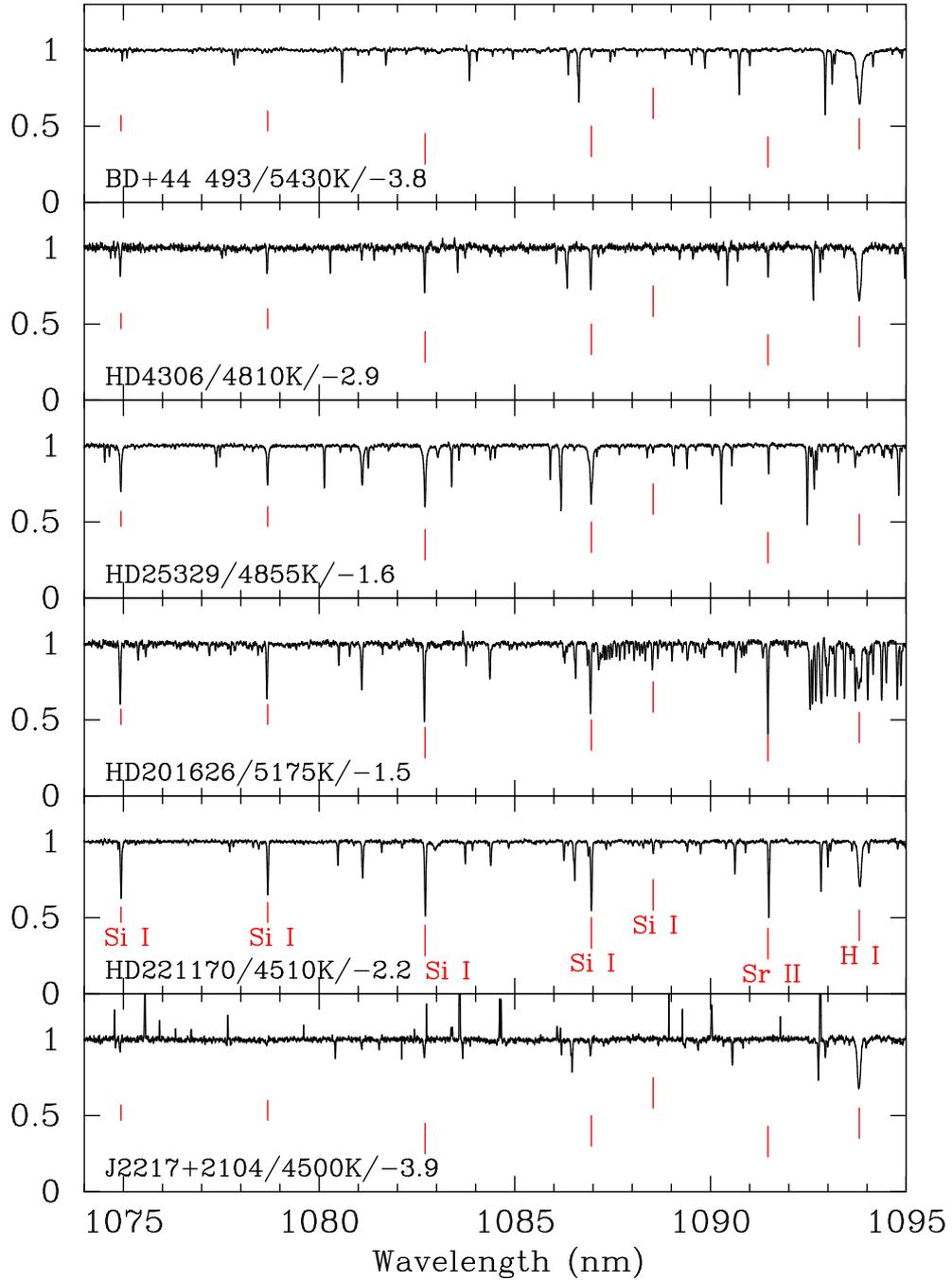}
   \end{center}
\caption{Examples of near-infrared spectra obtained with Subaru/IRD. The object name, the effective temperature and [Fe/H] value are presented in each panel. The spectral lines of Si and Sr used for the abundance analysis, as well as a hydrogen line, are marked by red vertical bars. In the CH star HD~201626, CN absorption bands are also found in 1087--1095~nm.}\label{fig:sp}
\end{figure}

The stellar parameters required for abundance analysis based on model atmospheres are taken from the literature and listed in table~\ref{tab:param}. In the most studies, the effective temperatures and the surface gravities are determined from colors (e.g., $V-K$) and assumption that the same Fe abundances are derived from neutral and ionized Fe lines, respectively. The errors of the parameters reported in the literature are typically 100~K for effective temperature, 0.3~dex for the surface gravity, 0.3~dex for [Fe/H] and 0.3~km~s$^{-1}$ for micro-turbulent velocity, or smaller.  
Examples of the spectra are shown in figure~\ref{fig:sp}.
The signal-to-noise (S/N) ratios of the spectra at 1050 and 1600~nm, which are estimated from photon counts, are given in table~\ref{tab:error}. 

\section{Abundance analysis}\label{sec:ana}
\subsection{Equivalent width measurements}\label{sec:ew}
Spectral line data for Si and the Sr triplet are taken from VALD \citep{Kupka1999AAS} and \citet{Grevesse2015AA}, respectively. The source of the Si line data in VALD is \citet{Kelleher2008JPCRD}. According to their evaluation, the accuracy of the transition probabilities of Si lines used in our analysis is B or C+, which corresponds to errors of 0.06~dex or better in $\log gf$ values. The line data, i.e., wavelengths, lower excitation potentials, and transition probabilities ($\log gf$ values) are listed in table~\ref{tab:ew}. Equivalent widths given in the table are obtained by fitting Gaussian profiles to the line profiles for the giant stars. For the spectrum of the main-sequence star HD~25329, in which spectral lines exhibit non-negligible wing components, Voigt profiles are fit to measure the equivalent widths. 

Errors of the equivalent widths ($\sigma_{W}$) are estimated at the wavelengths representing the $Y$-,$J$- and $H$-bands by the formula of \citet{Norris2001ApJ},  adopting $R=70,000$, $n_{\rm pix}=$10, and the S/N ratios given in table~\ref{tab:error}.   The $\sigma_{W}$ values range over 0.2--0.9~pm, depending on the data quality. These values are used to estimate the abundance errors due to spectral quality (see below).

\subsection{Si abundances}

Abundances of Si and Sr are determined by the standard LTE analysis using model atmospheres from the ATLAS/NEWODF grid \citep{Castelli2003IAUS} with enhancement of the $\alpha$ elements. Abundance analyses are made employing
a one-dimensional LTE spectral synthesis code that is based
on the same assumptions as the model atmosphere program of \citet{Tsuji1978AA}. 
The line broadening from the approximation of \citet{Unsold1955}, enhanced by a factor of 2.2, is adopted as done by \citet{Aoki2005ApJ}. We confirm that this treatment well reproduces the line profiles calculated with broadening parameters of \citet{Barklem2000AAS} for the lines for which the parameters are available. Our synthetic spectra well reproduce the Si lines in the solar spectrum for lines with equivalent widths smaller than 20~pm. For stronger lines, the line core profile is not well reproduced. This would be due to the non-LTE effect, as reported by \citet{Zhang2016ApJ} who studied the effect for Si lines in the $H$-band. The non-LTE effect is larger for stronger lines, and is not significant for weak lines found in metal-poor stars. The recent study by \citet{Masseron2021AA} for $H$-band lines reports that the non-LTE effect is dependent on spectral lines, but is smaller than 0.05~dex for metal-poor stars in globular clusters. We note that they also conducted 3D-LTE analysis and report significantly lower Si abundances. They conclude, however, that more extended self-consistent 3D–NLTE computations are required.

Stellar parameters in table~\ref{tab:param} are adopted from the literature with no modification, except for the micro-turbulent velocities of HD~221170 and HD~201626, which are determined by demanding that the derived Si abundances from individual lines do not exhibit a  dependence on the line strengths. For the other four stars, no useful constraint on the micro-turbulent velocity is obtained from Si lines, because the number of lines is too small or lines are very weak.  The Si abundances ($\log \epsilon$ values) determined from individual lines are listed in table~\ref{tab:abline}. 

\begin{figure}
 \begin{center}
   \includegraphics[width=8cm]{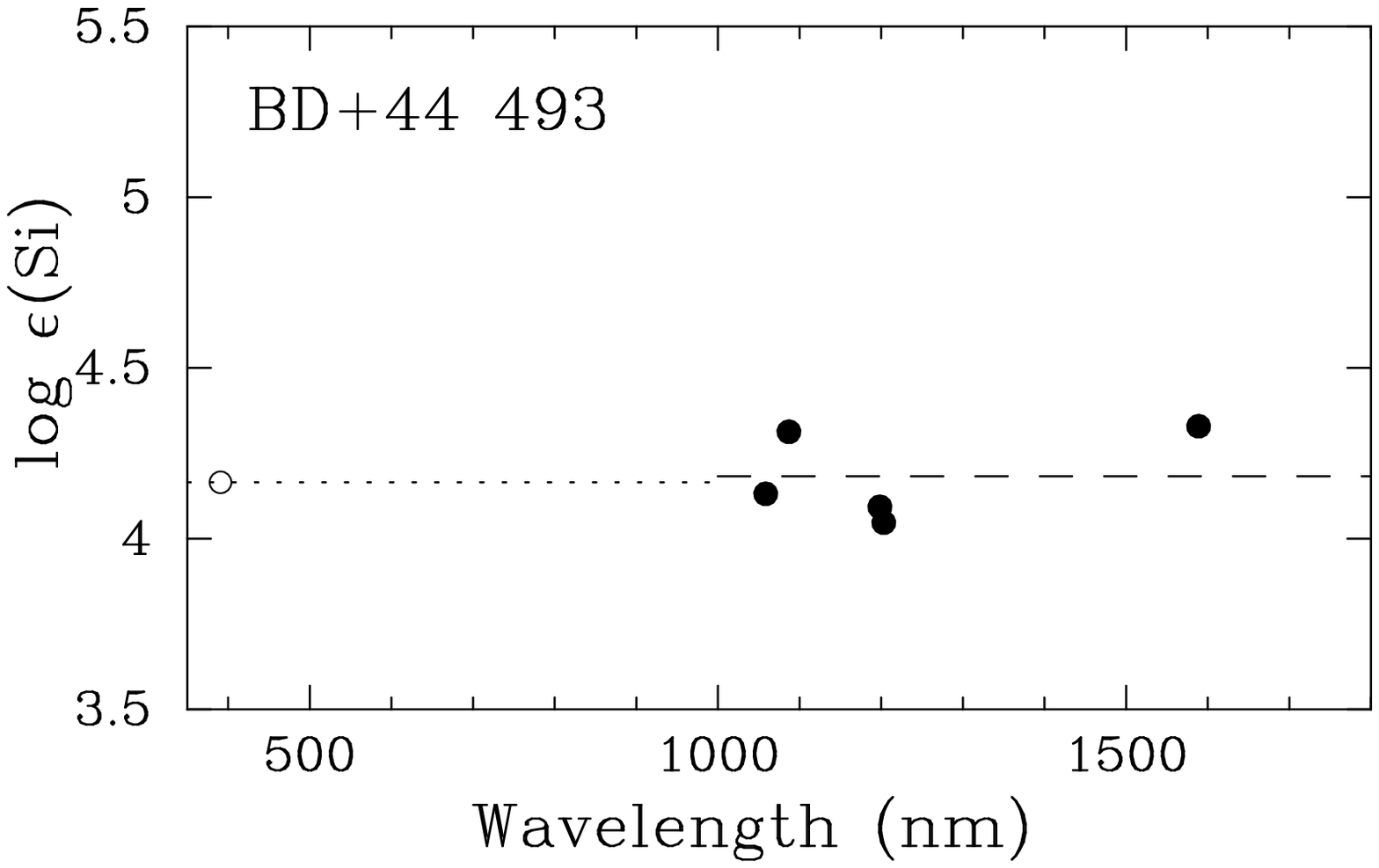}
   \includegraphics[width=8cm]{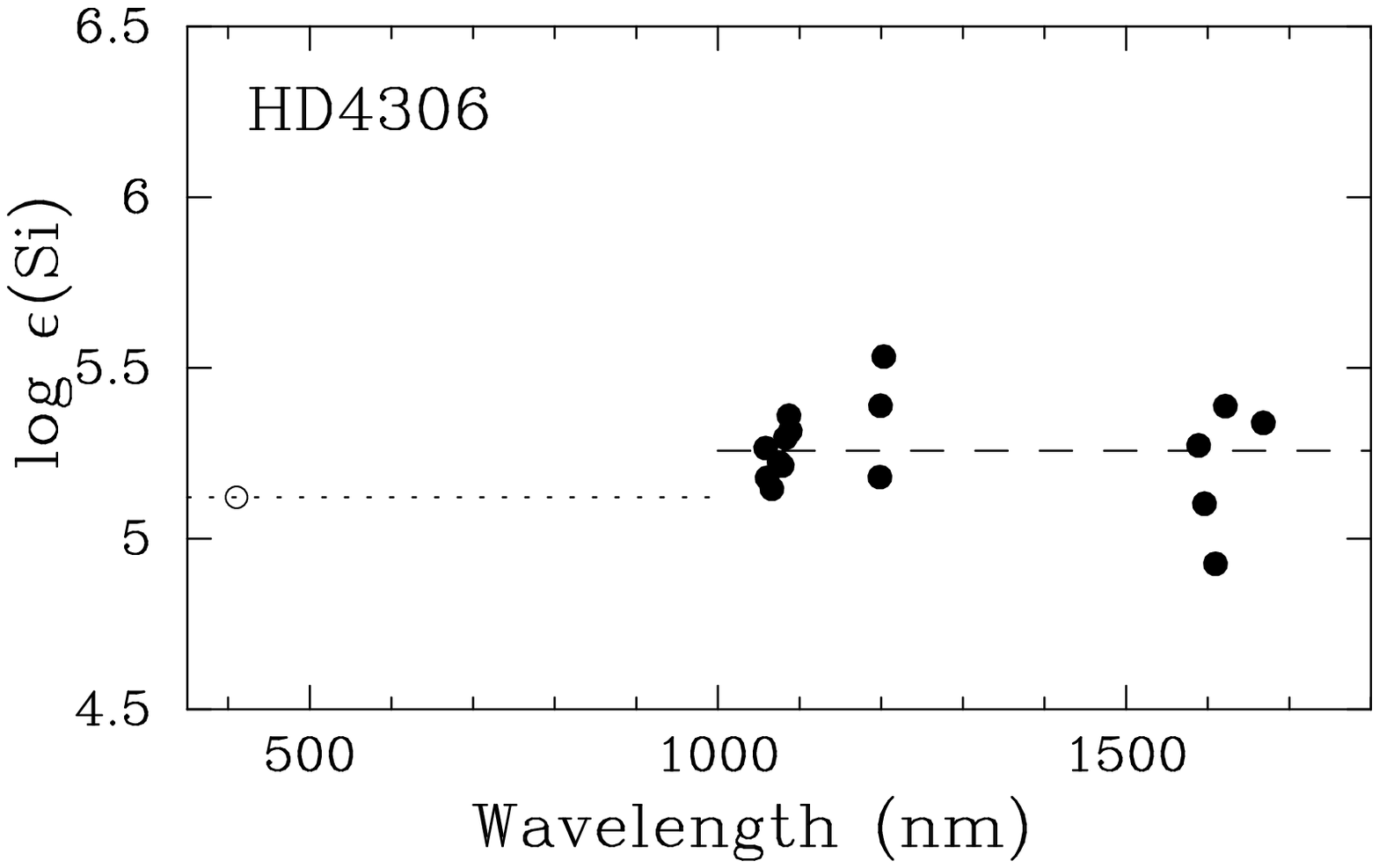}
   \includegraphics[width=8cm]{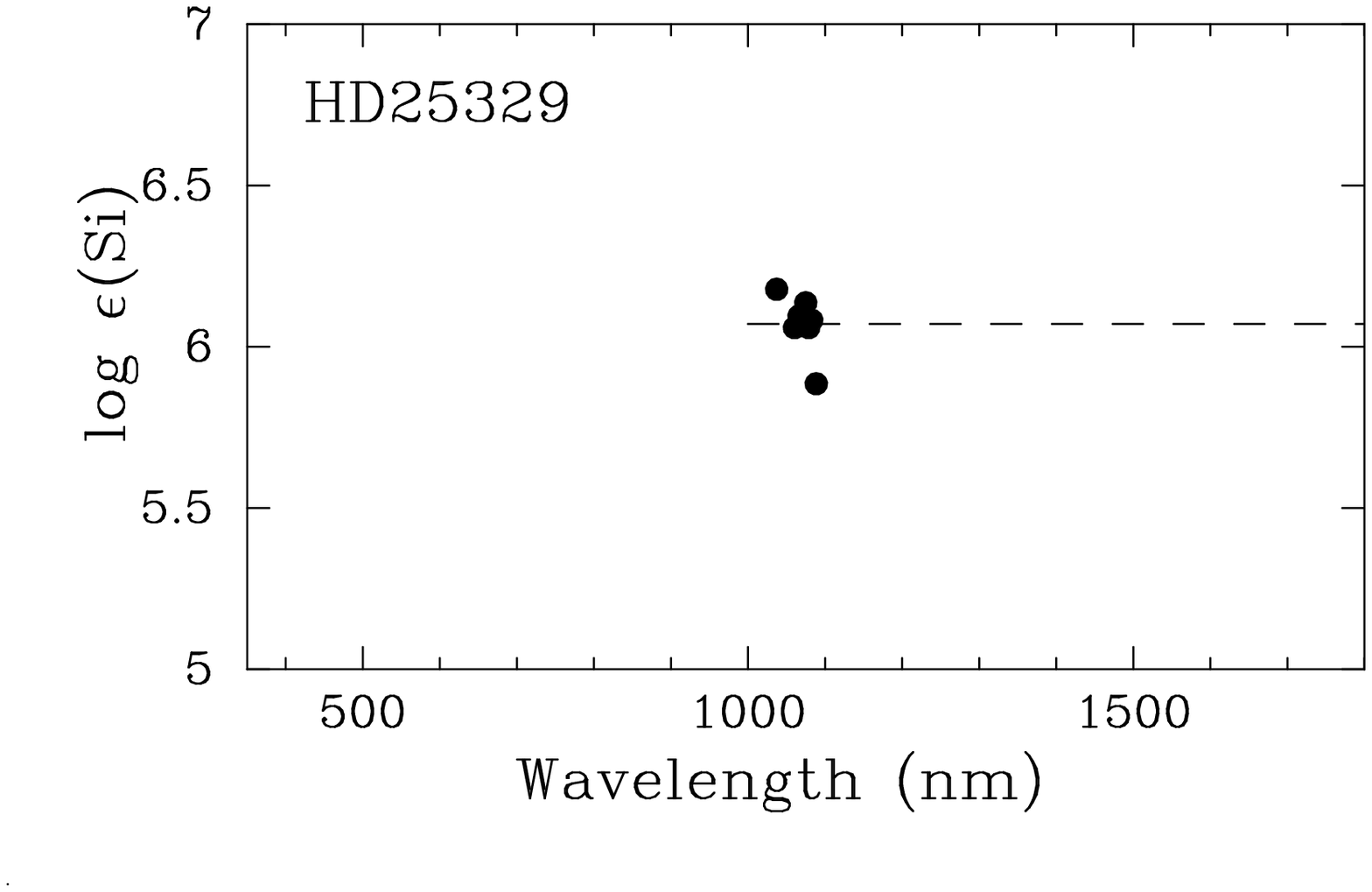}
   \includegraphics[width=8cm]{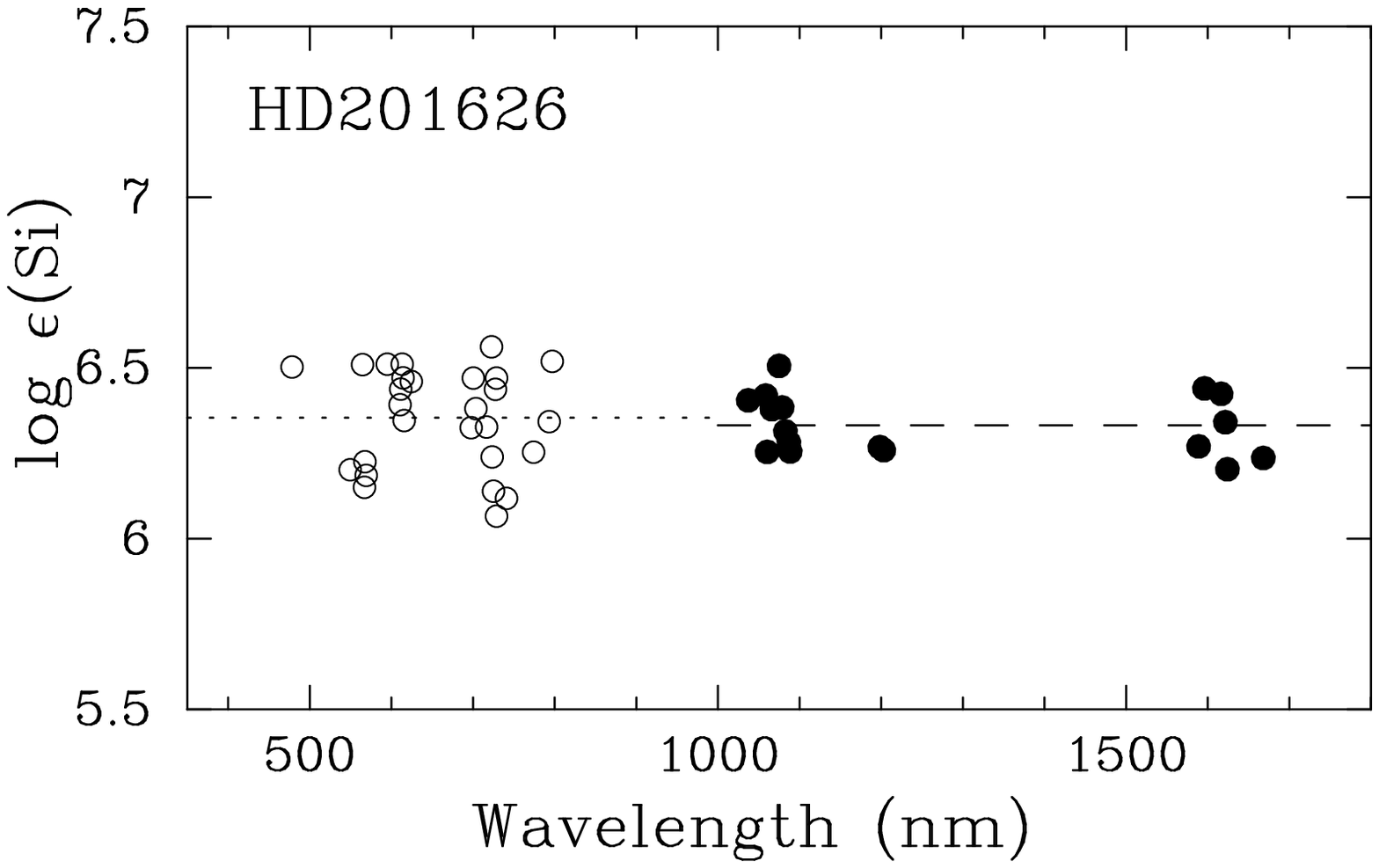} 
   \includegraphics[width=8cm]{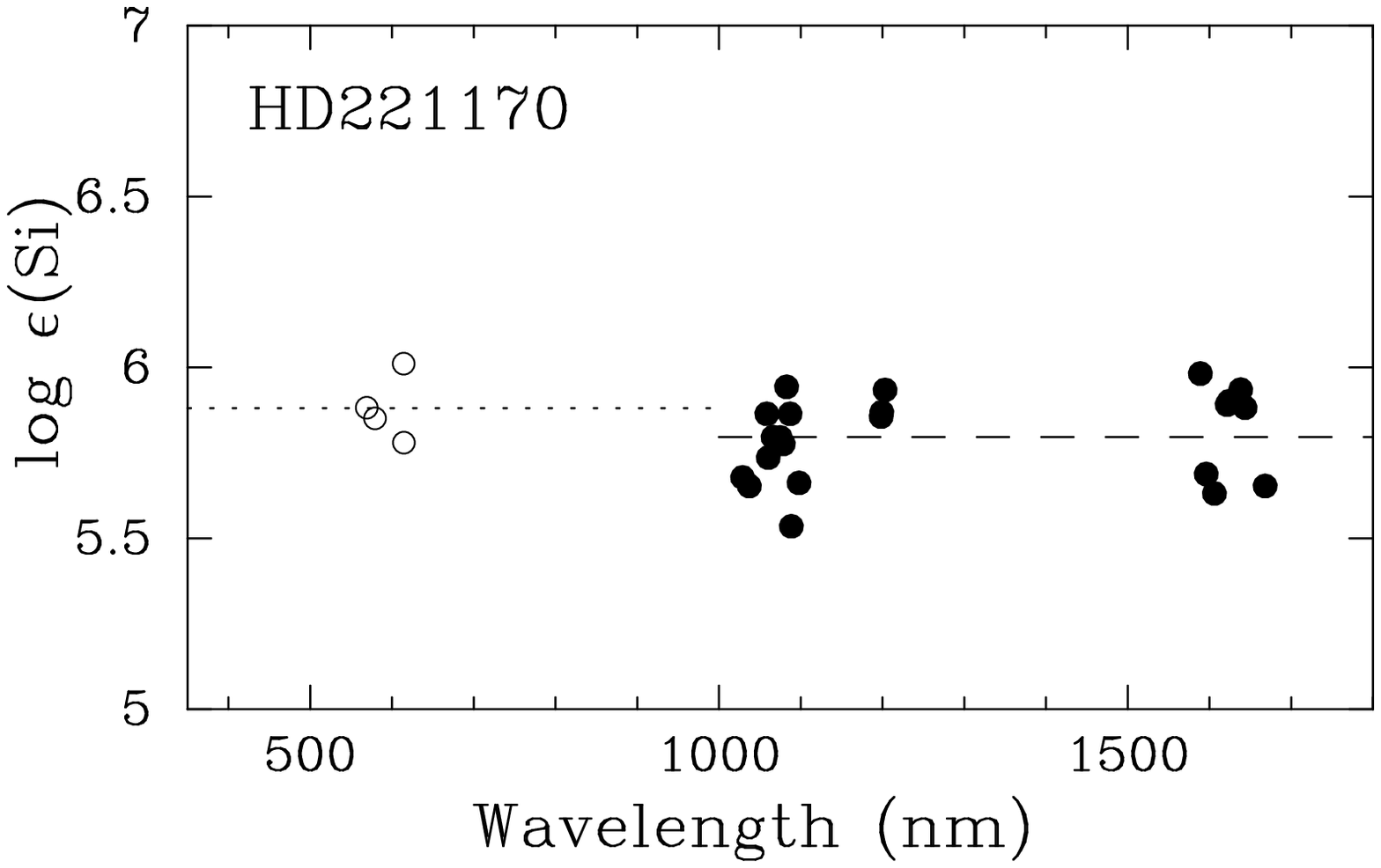}
   \includegraphics[width=8cm]{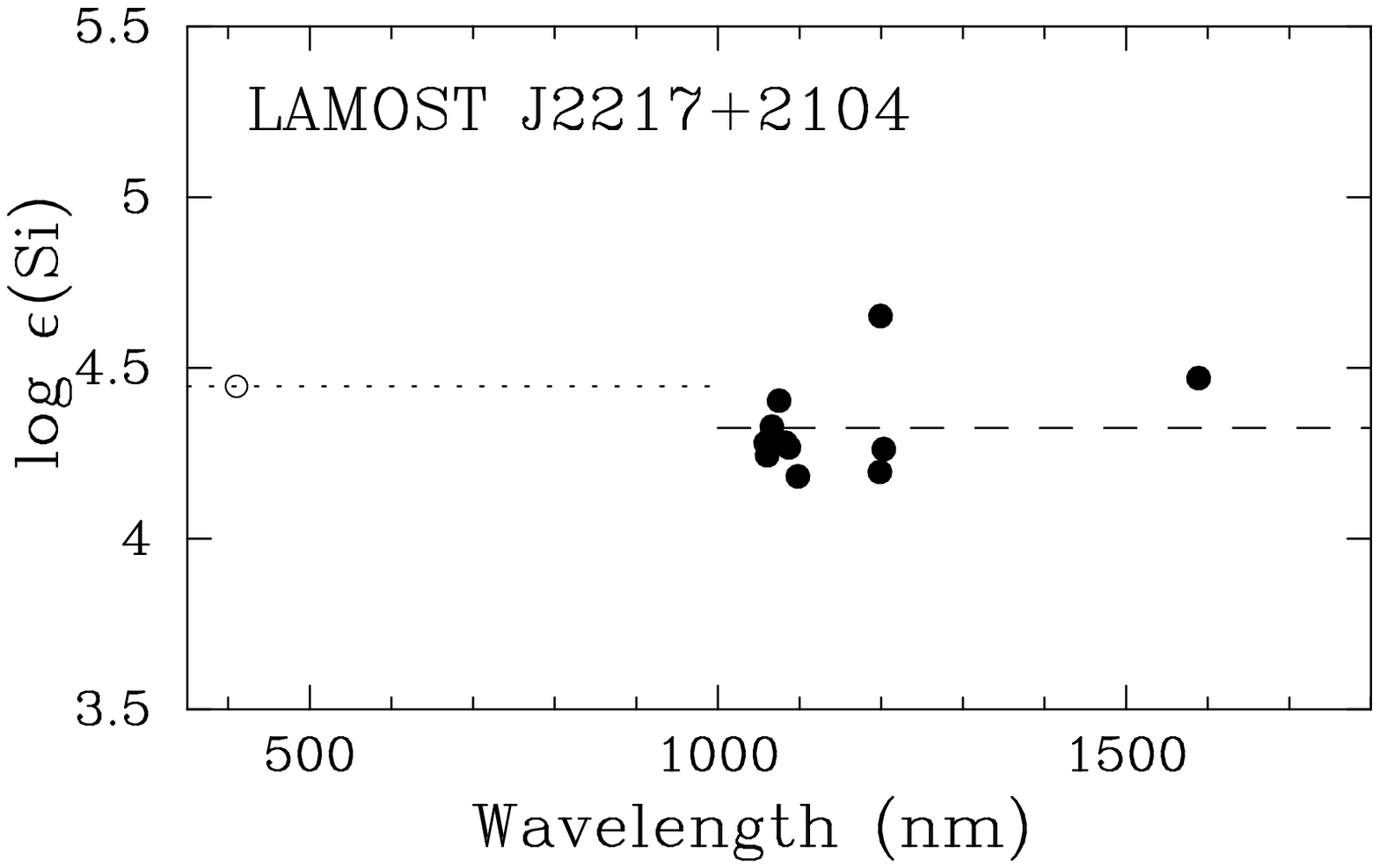}
 \end{center}
\caption{Si abundances determined from the near-infrared spectra (filled circles) in the present work. Open circles indicate the Si abundances determined using equivalent widths reported in the literature (see references in the text). Dotted and dashed lines show the averages of abundances determined from individual lines in the optical and near-infrared ranges, respectively. For HD~25329, the Si abundance from optical spectra is not available.}\label{fig:si1}
\end{figure}

Figure~\ref{fig:si1} shows the Si abundances derived from individual lines as a function of wavelengths. The Si abundances obtained from the optical lines by our analysis using equivalent widths in the literature (table~\ref{tab:param}) are also shown for comparison purposes. The equivalent widths of the optical lines of HD~201626 used in  \citet{Placco2015ApJ} are provided by Placco (private communication). This comparison demonstrates that the abundances derived from the optical and near-infrared spectra show good agreement. The abundance results from near-infrared spectra based on a much larger number of Si lines significantly improve the reliability of the derived abundances. For HD~201626, the Si abundance has been determined from many optical lines by \citet{Placco2015ApJ}. The result obtained from near-infrared spectra by the present work shows an excellent agreement. The scatter of the abundance results from near-infrared lines are slightly smaller than that from optical lines.

The number of Si lines measured for BD+44$^{\circ}$493 and HD~25329 is relatively small. This is because the  former is an extremely metal-poor ([Fe/H]$=-3.8$) subgiant star with very weak absorption lines in general. The latter is a relatively metal-rich ([Fe/H]$=-1.6$) cool main-sequence star, in which absorption features are broader and/or shallower than in red giants. LAMOST J~2217+2104 is an extremely metal-poor ([Fe/H]$=-3.9$) giant, but the Si lines are sufficiently strong due to its low temperature and relatively large excess of this element (see below).  

The Si abundances given in table~\ref{tab:si} are the average of those obtained from individual lines for each object. The abundances obtained from optical lines are included to obtain the $\log\epsilon$ values in this table, whereas the average values given in table~\ref{tab:abline} (bottom line) are those for near-infrared lines.

Table~\ref{tab:si} provides the standard deviation of the Si abundances derived from individual lines ($\sigma$) for each star. The $\sigma$ values are at the level of 0.1~dex, but larger in the spectra of stars with lower S/N ratios (i.e., HD~4306 and LAMOST~J~2217+2104). 

The line-by-line scatter should be primarily due to the errors of the equivalent widths and uncertainties of the spectral line data. 
We estimate the errors in the Si abundance measurements due to errors in equivalent widths by applying the analysis for equivalent widths changed by the $\sigma_{W}$ in table~\ref{tab:error}. The values in the $Y$-$J$-band and $H$-band are treated separately. The results are given in table~\ref{tab:error} as $\sigma_{\log\epsilon}$. The results depend on the size of $\sigma_{W}$ and the strengths of Si lines used in the analysis; the impact of  $\sigma_{W}$ is larger for weaker lines, whereas abundances derived from strong lines are sensitive to  the changes of equivalent widths due to saturation effects. The  $\sigma_{\log\epsilon}$ values are about 0.1~dex for HD~4306 and LAMOST~J~2217+2104, for which the  $\sigma_{W}$ values are relatively large due to relatively low S/N ratios. For other stars, the $\sigma_{\log\epsilon}$ values are much smaller than 0.1~dex. The value of HD~221170 is quite small (0.044~dex), even though the S/N ratio is only slightly better than for LAMOST~J~2217+2104. This is because the strengths of the Si lines used in the analysis in HD~221170 are moderate, whereas the Si lines are quite weak for LAMOST~J~2217+2104. 

Another source of errors in abundance measurements is the uncertainties of the spectral line data. 
We calculate the differences of abundances derived from individual lines from the mean abundance (bottom line of table~\ref{tab:abline}) for each star ($\delta_{i}=\log\epsilon_{i}-<\log\epsilon>$ for line $i$). Then we obtain the average of the abundance differences for each line, which is given in table~\ref{tab:abline} as $<\delta_{i}>$ for lines that are measured in more than two stars. Excluding the two lines at 1088~nm and 1199~nm, the deviations are smaller than 0.1~dex. We might correct the abundances from individual lines using these results. However, since they are based on at most five objects and still uncertain, we do not make corrections in the present work. Excluding the line at 1199~nm, which shows the largest deviation (0.177~dex), the standard deviation of the $<\delta_{i}>$ values is 0.06~dex. This value ($\sigma_{\rm line}$) is comparable to the uncertainties of the transition probabilities of Si lines (see \S~\ref{sec:ew}).


The $\sigma$ value given in table~\ref{tab:si} is mostly explained by $\sigma_{\log\epsilon}$ and $\sigma_{\rm line}$. The random error of the abundance measurement is given as $\sigma N^{-1/2}$.

We estimate the errors due to uncertainties of stellar parameters from the abundance changes for changes of the stellar parameters, $\Delta T_{\rm eff}=100$~K, $\Delta\log g=0.3$~dex, $\Delta$[Fe/H]$=0.3$~dex, and $\Delta v_{\rm turb}=0.5$~km~s$^{-1}$ for HD~221170. The quadrature sum of the changes is 0.13~dex, which is dominated by the changes of micro-turbulent velocity and effective temperature. This value, $\sigma_{\rm param}$, and the random error obtained above are added in quadrature to derive total errors, which is shown in the top panel of  figure~\ref{fig:absi}.  

\subsection{Sr abundances}

Strontium abundances are determined for four stars, and listed in table~\ref{tab:sr}. The errors are estimated by  $\sigma N^{-1/2}$, as for the Si abundance, although they are uncertain because the number of lines used in the analysis is small. None of the triplet lines are detected in the spectra of the two extremely metal-poor stars BD+44$^{\circ}$493 and LAMOST~J~2217+2104. 
For HD~4306, the equivalent width of the Sr II 407.8~nm line provided by \citet{Honda2004ApJS} is included in our analysis.

The Sr abundances of HD~4306 and HD~221170 obtained from the near-infrared spectra are 0.2--0.3~dex higher than the values obtained from optical spectra by \citet{Honda2004ApJ} ($\log \epsilon$(Si)$=-0.08$ from the Sr II 407.8~nm line) and \citet{Ivans2006ApJ} ($\log \epsilon$(Si)=0.85 obtained from four lines, including very weak ones). The Sr abundance of HD~4306 derived by our analysis of the Sr II 407.8~nm line is even lower ($\log\epsilon$(Si)$=-0.15$): the small difference from the result of \citet{Honda2004ApJ} could be due to the difference in the treatment of the line broadening. The solar Sr abundance derived from the triplet lines is reported in \citet{Grevesse2015AA} for the cases with different assumptions. Their result obtained by the 1D-LTE analysis based on the model atmosphere of \citet{Holweger1974SoPh} does not show any discrepancy from the final result for the solar Sr abundance including 3D and NLTE. Our LTE analysis of the solar Sr abundance using the same line list and the model atmosphere of \citet{Holweger1974SoPh} reproduces their result within 0.1~dex, confirming the consistency of our analysis. Further examination of the possible discrepancy (at the 0.2-0.3~dex level) of Sr abundances between the blue resonance doublet lines and near-infrared triplet lines is required  from near-infrared spectra for a larger sample.

\section{Discussion}\label{sec:disc}

\begin{figure}
 \begin{center}
   \includegraphics[width=13cm]{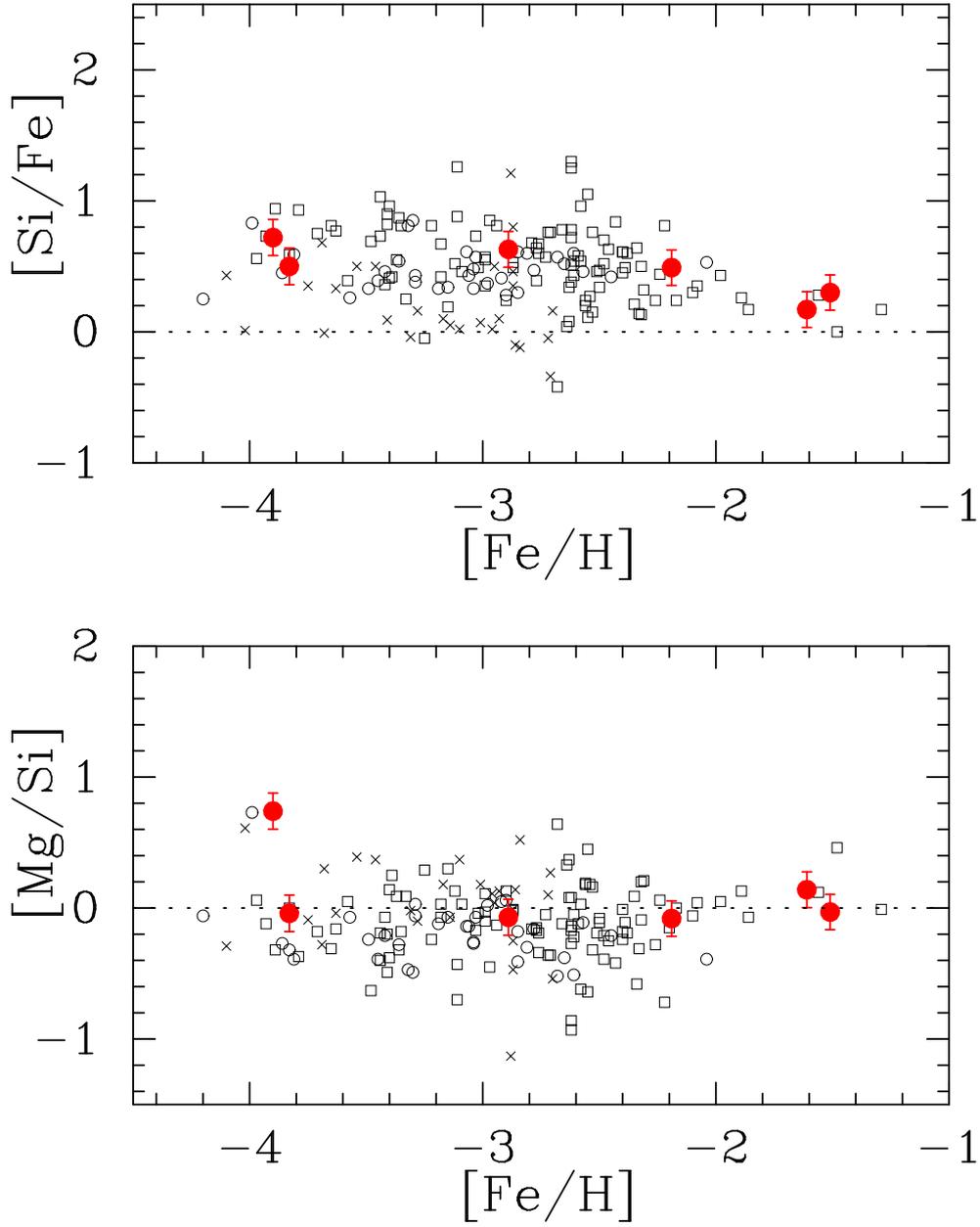}
 \end{center}
\caption{[Si/Fe] and [Mg/Si] as a function of [Fe/H]. The results obtained by the present work are shown by filled circles. The results obtained by previous studies based on optical spectra are shown by open circles \citep{Cayrel2004AA}, crosses \citep{Yong2013ApJ}, and open squares \citep{Jacobson2015ApJ}.}\label{fig:absi}
\end{figure}

Figure~\ref{fig:absi} (upper panel) shows [Si/Fe] as a function of [Fe/H] for our sample. The results of three previous studies \citep{Cayrel2004AA, Yong2013ApJ, Jacobson2015ApJ} based on measurements of optical lines are also shown for comparison purposes. All the six stars show a clear over-abundance of Si, as expected for very metal-poor stars. The over-abundance of HD~25329,  [Si/Fe]$=+0.17$, is relatively small. At the metallicity of this star ([Fe/H]$=-1.6$) the over-abundances of the $\alpha$-elements of some halo stars are smaller than those found for VMP stars. These stars could have been accreted from small stellar systems that have experienced chemical evolution with longer timescales, resulting in lower abundances of the $\alpha$-elements due to contributions from type Ia supernovae. We note that the Si abundance of HD~201626, with [Fe/H]$=-1.5$, is also lower ([Si/Fe]$=+0.30$) than those of the four VMP stars.



The [Si/Fe] values determined from the near-infrared spectra for the other four stars are almost constant with little scatter. The [Si/Fe] values derived from optical spectra show larger scatter than those of Mg, which might be caused by measurement errors. Further studies for abundance trends and scatter of [Si/Fe] in VMP stars based on near-infrared spectra are obviously required for a larger sample to obtain definitive results on the [Si/Fe] abundance distributions.  

The [Mg/Si] abundance ratios are shown in the lower panel of figure~\ref{fig:absi}. The [Mg/Si] values for four stars, including HD~25329 and HD~201626, agree with the solar level (i.e., [Mg/Si]$=0$) within the measurement errors. The [Mg/Si] of LAMOST J~2217+2104 is high, reflecting the large excess of Mg in this star  \citep{Aoki2018PASJ}. This object is a CEMP-no star with a large excess of Mg.  The Si is also over-abundant, but is not as much as for Mg. An interpretation of this peculiar abundance ratios is a larger-scale mixing and fallback, resulting in smaller amount of ejecta from the Si layer and inside \citep{Ishigaki2018ApJ}. Excluding this object, the scatter of [Mg/Si] is very small. The apparently quite low scatter requires confirmation from measurements for a larger sample of metal-poor stars.

[Sr/Fe] values of HD~25239 and HD~221170 follow the trend found in halo stars ($0.0<$[Sr/Fe]$<0.5$) in very metal-poor stars. The value of HD~4306 is within the wide distribution of [Sr/Fe] found in extremely metal-poor stars. The CH star HD~201626 exhibits a clear excess of Sr, whereas it is smaller than found for heavier neutron-capture elements, e.g., Ba \citep{Placco2015ApJ}. This is an anticipated result from the s-process models for very low-metallicity, where heavier elements are more enhanced due to the higher neutron exposures to smaller amount of seed nuclei.

The errors of equivalent widths ($\sigma_{W}$) at around 1~$\mu$m are about 0.4~pm (table~\ref{tab:error}) for a spectrum with $S/N\sim$100. If 3$\sigma_{W}$ is adopted as an upper limit, the detection limit of the Si abundance is [Si/H]$\sim -4.5$ and $-4.0$ for red giants ($T_{\rm eff}\sim 5000~K$) and subgiants ($T_{\rm eff}\sim 5500~K$), respectively. This indicates that future measurements of Si abundances for VMP stars from high-resolution near-infrared spectra are very promising, and the abundance trends and scatter of [Si/Fe] will be well-determined. 

The detection limit of the Sr abundance from the near-infrared triplet lines estimated by the same assumption is [Sr/H]$\sim -3.6$ and $-2.9$ for red giants  and subgiants, respectively. This means that these lines are not sufficiently strong for abundance measurements of Sr in EMP stars if Sr is under-abundant. Indeed, no Sr line is detected in our spectra of BD~+44$^{\circ}$493 and LAMOST~J~2217+2104. Instead, these lines are useful to determine precise abundances with relatively high Sr abundances, in which the resonance lines in the blue range are too strong and/or severely affected by blending of other lines. Hence, the near-infrared triplet lines and the blue resonance lines are complementary to cover the wide ranges of Sr abundance ratios in VMP and EMP stars. 

The  scatter of [Sr/Fe] found in metal-poor stars is as large as 3~dex (e.g., \cite{McWilliam1995AJ,Honda2004ApJ}). This is much larger than the measurement errors from the resonance lines. Improving the abundance measurements using the triplet lines will contribute to determine more detailed distributions of these abundance ratios, which may identify some fine structure or clustering in the abundance distributions (e.g., \cite{Roederer2013AJ,Aoki2020AA}).

\section{Summary and concluding remarks} 

We have determined Si and Sr abundances for six metal-poor stars from measurements of spectral lines identified in high-resolution near-infrared spectra obtained with the Subaru Telescope InfraRed Doppler instrument (IRD). The Si abundances derived  from infrared spectra exhibit clear trends and over-abundances. Further measurements of the near-infrared lines will provide reliable Si abundances to determine the abundance trends and scatter, which can be used to place strong constraints on chemical-evolution models. The Sr triplet lines in the $J$-band are also useful for determining the abundance distribution of this element in metal-poor stars covering objects with high Sr abundances. 

\begin{ack}

This research is based on data collected at Subaru Telescope, which is operated by the National Astronomical Observatory of Japan. We are honored and grateful for the opportunity of observing the Universe from Maunakea, which has the cultural, historical and natural significance in Hawaii.
This work was supported in part by Strategic International Research Exchange Promotion Program of the National Institutes of Natural Sciences (NINS) and 
the National Science Foundation under Grant No. OISE-1927130 (IReNA). 
W.A. is supported by JSPS KAKENHI grant No. 21H00055.
T.C.B. acknowledges partial support from grant PHY 14-30152
(Physics Frontier Center/JINA-CEE), awarded by the U.S.
National Science Foundation. M.T. is supported by JSPS KAKENHI grant Nos.18H05442, 15H02063, and 22000005. The work of V.M.P. is supported by NOIRLab, which is managed by the Association of Universities for Research in Astronomy (AURA) under a cooperative agreement with the National Science Foundation.

\end{ack}

\clearpage

\clearpage
\begin{table}
  \tbl{Stellar Parameters}{%
  \begin{tabular}{lccccl}
      \hline
    Star & $T_{\rm eff}$ & $\log g$ & [Fe/H] & $v_{\rm turb}$ & Reference \\
           & (K)  & (cgs) &  & (km~s$^{-1}$) & \\
      \hline
BD+44$^{\circ}$493  & 5400 & 3.4  &  $-3.83$ &  1.3  &  \citet{Ito2013ApJ}\\
HD~4306     & 4810 & 1.8  &  $-2.89$ &  1.6  &   \citet{Honda2004ApJ} \\
HD~25329    & 4855 & 4.73 &  $-1.61$ &  1.63 &  \citet{Luck2017AJ}  \\
HD~201626   & 5175 & 2.80 &  $-1.51$  & 1.8$^{*}$ &   \citet{Placco2015ApJ} \\
HD~221170   & 4510 & 1.0  &  $-2.19$ &  1.8  &   \citet{Ivans2006ApJ} \\
LAMOST~J~2217+2104 & 4500 & 0.9  &  $-3.90$ &  2.3$^{*}$ &  \citet{Aoki2018PASJ} \\
      \hline
    \end{tabular}}\label{tab:param}
\begin{tabnote}
\footnotemark[$*$] Micro-turbulent velocities updated by the present work.  \\ 
\end{tabnote}
\end{table}

\begin{table}
  \tbl{Errors of Equivalent Width and Abundance}{%
  \begin{tabular}{lrrcrrcc}
      \hline
Star    & \multicolumn{2}{c}{1050~nm} &  & \multicolumn{2}{c}{1600~nm} & & $\sigma_{\log\epsilon}$ \\
\cline{2-3} \cline{5-6} 
          & S/N & $\sigma_{W}$ (pm)                 &  & S/N & $\sigma_{W}$ (pm) & & \\
      \hline
BD+44$^{\circ}$493 & 187 & 0.229 &  & 253 & 0.271 &  & 0.067 \\
HD~4306    & 57  & 0.758 &  &  78 & 0.878 &  & 0.110 \\
HD~25329   & 130 & 0.329 &  & 207 & 0.331 &  & 0.020 \\
HD~221170  & 134 & 0.319 &  & 230 & 0.298 &  & 0.033 \\
HD~201626  & 92  & 0.465 &  & 145 & 0.473 &  & 0.044 \\
LAMOST~J~2217+2104 & 73 & 0.589 &  & 110 & 0.626 &  & 0.107 \\
     \hline
    \end{tabular}}\label{tab:error}
\end{table}


\begin{table}
  \tbl{Equivalent Widths}{%
  \begin{tabular}{lrrrrrrrr}
      \hline
Wavelength & L.E.P. & $\log gf$ & BD+44$^{\circ}$493 & HD~4306 & HD~25329 & HD~201626 & HD~221170 & LAMOST~J~2217+2104 \\
(nm)     & (eV)     &       &       &       &       &       &    & \\
      \hline
      \multicolumn{9}{c}{Si {\small I}} \\
      \hline
1028.894 & 4.920 &-1.511&       &       &       &       &  2.31 & \\
1037.126 & 4.930 &-0.705&       &       &  5.88 &  9.44 &  7.65 &  \\
1058.514 & 4.954 & 0.012&  0.79 & 7.73  &       & 16.11 & 15.88 & 2.87\\
1060.343 & 4.930 &-0.305&       & 5.20  & 11.38 & 11.70 & 12.23 & 1.56\\
1066.097 & 4.920 &-0.266&       & 5.37  & 11.99 & 13.32 & 13.30 & 2.07\\
1074.938 & 4.930 &-0.205&       & 6.29  & 13.76 & 15.22 & 13.85 & 2.65\\
1078.685 & 4.930 &-0.303&       & 5.59  & 11.45 & 13.12 & 12.84 &  \\
1082.709 & 4.954 & 0.302&       & 10.22 & 23.66 & 18.84 & 19.65 & 4.84\\
1086.954 & 5.082 & 0.371&  1.92 & 10.18 &       & 17.82 & 18.17 & 4.16\\
1088.533 & 6.181 & 0.221&       &  1.56 &  1.60 &  5.96 &  2.88 & \\
1097.931 & 4.954 &-0.524&       &       &       &       &  9.76 & 0.90\\
1198.420 & 4.930 & 0.239&  1.64 & 10.71 &       & 20.62 & 21.02 & 4.98\\
1199.157 & 4.920 &-0.109&       &  9.73 &       &       & 17.80 & 5.97\\
1203.150 & 4.954 & 0.477&  2.30 & 15.57 &       & 24.30 & 24.30 & 7.36\\
1588.844 & 5.082 & 0.000&  3.26 & 18.31 &       & 31.81 & 34.31 & 11.18\\
1596.008 & 5.984 & 0.200&       &  6.30 &       & 23.24 & 16.77 &  \\
1606.002 & 5.954 &-0.430&       &       &       &       &  7.34 & \\
1609.480 & 5.964 &-0.080&       &  2.94 &       &       &       &  \\
1616.371 & 5.954 &-0.850&       &       &       &  7.50 &       &  \\
1621.569 & 5.954 &-0.580&       &  2.85 &       &  9.93 &  8.99 &  \\
1624.185 & 5.964 &-0.760&       &       &       &  6.04 &  6.74 &  \\
1638.155 & 5.964 &-0.390&       &       &       &       & 12.52 &  \\
1668.077 & 5.964 &-1.060&       &  6.20 &       & 15.42 & 12.48 & \\
      \hline
      \multicolumn{9}{c}{Sr {\small II}} \\
      \hline
1003.665 & 1.810 &-1.202&       &  1.30 &  1.31 & 12.34 &  8.65 & \\ 
1032.731 & 1.840 &-0.248&       &  6.96 &  6.51 & 21.49 & 18.49 &  \\
1091.489 & 1.810 &-0.478&       &  5.46 &       & 19.77 & 16.89 & \\
\hline
    \end{tabular}}\label{tab:ew}
\end{table}

\begin{table}
  \tbl{Si Abundances derived from Individual Lines}{%
  \begin{tabular}{lrrrrrrr}
      \hline
Wavelength & BD+44$^{\circ}$493 & HD~4306 & HD~25329 & HD~201626 & HD~221170 & LAMOST~J~2217+2104 &  $<\delta_{i}>^{*}$ \\
(nm)     &       &       &       &       &       &       &     \\      \hline
1028.894 &       &       &       &       & 5.678 &        &  \\
1037.126 &       &       & 6.178 & 6.365 & 5.653 &        &  0.010\\
1058.514 & 4.131 & 5.265 &       & 6.372 & 5.864 & 4.280  &  0.010\\
1060.343 &       & 5.178 & 6.059 & 6.206 & 5.736 & 4.244  & -0.065\\
1066.097 &       & 5.146 & 6.095 & 6.329 & 5.796 & 4.328  & -0.011\\
1074.938 &       & 5.223 & 6.136 & 6.457 & 5.795 & 4.404  &  0.053\\
1078.685 &       & 5.214 & 6.059 & 6.334 & 5.776 &        & -0.010\\
1082.709 &       & 5.295 & 6.082 & 6.272 & 5.943 & 4.281  &  0.025\\
1086.954 & 4.313 & 5.360 &       & 6.238 & 5.864 & 4.267  &  0.036\\
1088.533 &       & 5.315 & 5.885 & 6.240 & 5.535 &        & -0.112\\
1097.931 &       &       &       &       & 5.662 & 4.182  &  \\
1198.420 & 4.093 & 5.180 &       & 6.227 & 5.856 & 4.195  & -0.062\\
1199.157 &       & 5.389 &       &       & 5.869 & 4.652  &  0.177\\
1203.150 & 4.047 & 5.533 &       & 6.225 & 5.934 & 4.262  &  0.028\\
1588.844 & 4.329 & 5.273 &       & 6.234 & 5.982 & 4.470  &  0.086\\
1596.008 &       & 5.102 &       & 6.410 & 5.688 &        & -0.051\\
1606.002 &       &       &       &       & 5.631 &        &  \\
1609.480 &       & 4.926 &       &       &       &        &  \\
1616.371 &       &       &       & 6.414 &       &        &  \\
1621.569 &       & 5.388 &       & 6.327 & 5.891 &        &  0.084\\
1624.185 &       &       &       & 6.196 & 5.901 &        &   \\
1638.155 &       &       &       &       & 5.935 &        &   \\
1668.077 &       & 5.339 &       & 6.213 & 5.653 &        & -0.049 \\
      \hline
mean  & 4.183 & 5.258 & 6.071 & 6.298 & 5.792 & 4.324     &  \\
     \hline
    \end{tabular}}\label{tab:abline}
\begin{tabnote}
\footnotemark[$*$] Average of abundance deviation from the average of the abundance for each star.  \\ 
\end{tabnote}
\end{table}


\begin{table}
  \tbl{Si Abundances}{
  \begin{tabular}{lccccccccc}
      \hline
    Object & \multicolumn{2}{c}{$\log\epsilon$(Si)} & $\log\epsilon$(Si) & [Si/Fe]$^{1}$ & $N$ & $\sigma$ & $\sigma N^{-1/2}$ & [Fe/H]$^{2}$ & [Mg/Fe]$^{2}$  \\
\cline{2-3} 
& optical  & NIR & all &  & & (dex) & (dex)& & \\
      \hline
BD+44$^{\circ}$493    & 4.165 & 4.183 & 4.180 &  0.50 &     6 & 0.106 & 0.043 & -3.83 & 0.46\\
HD~4306       & 5.121 & 5.258 &  5.250 &  0.63 &    17 & 0.135 & 0.033 & -2.89 & 0.56    \\
HD~25329      & ... & 6.071  &  6.071 &  0.17 &     7 & 0.086 & 0.033 & -1.61 & 0.31    \\
HD~201626     & 6.350 & 6.298 &  6.330 &  0.33 &    44 & 0.124 & 0.019 & -1.51  & 0.27 \\
HD~221170     & 5.881 & 5.798 &  5.810 &  0.49 &    26 & 0.120 & 0.024 & -2.19 & 0.41    \\
J2217+2104    & 4.446 & 4.324 &  4.334 &  0.72 &    12 & 0.130 & 0.038 & -3.90 & 1.46\\
      \hline
    \end{tabular}}\label{tab:si}
\begin{tabnote}
  \footnotemark[$1$] The solar abundance of $\log\epsilon_{\odot}$(Si)$=7.51$ is adopted.  \\
  \footnotemark[$2$] Taken from the literature given in table~\ref{tab:param}  \\ 
\end{tabnote}
\end{table}

\begin{table}
  \tbl{Sr Abundances}{%
  \begin{tabular}{lcccc}
      \hline
    Star & $\log\epsilon$(Sr) & [Sr/Fe]$^{1}$ & $N$ & $\sigma$(dex)   \\
      \hline
HD~4306       &  0.134  &  0.154  & 4 & 0.077 \\
HD~25329      &   1.436 &   0.176 &  2 & ... \\
HD~201626     &   2.576 &   1.216 &  3 & 0.025 \\
HD~221170     &   1.133 &   0.453 &  3 & 0.105  \\
      \hline
    \end{tabular}}\label{tab:sr}
\begin{tabnote}
  \footnotemark[$1$] The solar abundance of $\log\epsilon_{\odot}$(Sr)$=2.87$ is adopted.  \\
\end{tabnote}
\end{table}

\clearpage


\end{document}